\documentclass[prl,10pt,superscriptaddress,twocolumn]{revtex4}

\usepackage{graphicx}
\usepackage{amsmath}
\usepackage{color}
\usepackage{units}
\usepackage[latin1]{inputenc}
\usepackage{verbatim}
\usepackage{float}
\usepackage[english]{babel}
\usepackage{booktabs}	
\usepackage{epstopdf}
\usepackage{pgf}
\usepackage{pgfplots}
\usepackage{placeins}

\begin{document}


\title{Einstein-Podolsky-Rosen\,--\,entangled motion of two massive objects}

\author{Roman Schnabel}
\address{Institut f\"ur Laserphysik and Zentrum f\"ur Optische Quantentechnologien, Universit\"at Hamburg, Luruper Chaussee~149, 22761 Hamburg, Germany}
\address{Institut f\"ur Gravitationsphysik, Leibniz Universit\"at Hannover  and Max-Planck-Institut f\"ur Gravitationsphysik
(Albert-Einstein-Institut),\\ Callinstra{\ss}e~38, 30167 Hannover, Germany}
\email[corresponding author:\\]{roman.schnabel@physnet.uni-hamburg.de}

\begin{abstract}
In 1935, Einstein, Podolsky and Rosen (EPR) considered two particles in an entangled state of motion to illustrate why they questioned the completeness of quantum theory. 
In the past decades, microscopic systems with entanglement in various degrees of freedom have successfully been generated, representing compelling evidence to support the completeness of quantum theory. Today, the generation of an EPR-entangled state of motion of two \emph{massive} objects of up to the kilogram-scale seems feasible with state-of-the-art technology. Recently, the generation and verification of EPR-entangled mirror motion in interferometric gravitational wave detectors was proposed, with the aim of testing quantum theory in the regime of macroscopic objects, and to make available nonclassical probe systems for future tests of modified quantum theories that include (non-relativistic) gravity. 
The work presented here builds on these earlier results and proposes a specific Michelson interferometer that includes two high-quality laser mirrors of about 0.1\,kg mass each. 
The mirrors are individually suspended as pendula and located close to each other, and cooled to about 4\,K. 
The physical concepts for the generation of the EPR-entangled centre of mass motion of these two mirrors are described. 
Apart from a test of quantum mechanics in the macroscopic world, the setup is envisioned to test predictions of yet-to-be-elaborated modified quantum theories that include gravitational effects.  
\end{abstract}

\maketitle

\section{Introduction}
Quantum theory has been found to be in full agreement with all experimental tests so far. Nevertheless, modified quantum theories are under development, with the goal of solving the so-called measurement problem \cite{Bassi2003}, and providing potential bridges towards a not-yet-found unified theory of quantum gravity \cite{Kiefer2006}. 
Solid-state mechanical systems in pure nonclassical quantum states of motion, such as squeezed states and entangled states of motion, constitute interesting probe systems to test distinct predictions from modified quantum theories. A prominent example is the effect of gravity decoherence \cite{Diosi1987,Diosi1989,Penrose1996, Diosi1998}, where the system's own gravitational field is conjectured to introduce decoherence to the system's motional state. An experimental test in which the motional degrees of freedom of a micromechanical oscillator are entangled with the presence or absence of a single photon  was proposed in \cite{Marshall2003}.  
More recently, it was proposed to test predictions of an effective Schr\"odinger-Newton equation for the center of mass motion of a massive object \cite{Yang2013}. Here, much heavier mechanical oscillators with eigenfrequencies in the audio-band (the lower the better) are preferred.

One might expect that further tests of macroscopic quantum mechanics will be proposed that call for heavy masses in nonclassical states of motion. Pendulum-suspended massive objects might be appropriate systems.
They could constitute interesting probes of modified quantum theories because, at Fourier frequencies above the pendulum eigenfrequency, they act as quasi-free falling test masses of space-time. This property is used in gravitational wave detectors \cite{Harry2010}. 
Unfortunately, pendulum-suspended mirrors usually have a huge thermal excitation and are far from being in a pure quantum mechanical state.

In a recent series of fascinating experiments, mesoscopic mechanical oscillators of micrometre sizes and GHz resonance frequencies were cooled into their quantum ground states \cite{OConnell2010,Teufel2011,Chan2011}. The fluctuation of the oscillator's position was thus of the order of the zero-point fluctuation  
$\Delta \hat x_{\rm{zpf}} =  \sqrt{\hbar/(2 m \Omega)}$,  where $m$ is the motional mass of the localised acoustic mode,  $\Omega/2\pi$ its resonance frequency, and $\hbar$ the reduced Planck constant. In all three experiments the average thermal phonon excitation of the oscillator was below one. At a given temperature $T$, the thermal occupation of a mechanical oscillator according to the Bose-Einstein statistics is given by $\bar{n} = (e^{\hbar \Omega/kT}-1)^{-1}$, with $k$ the Boltzmann constant. 
As a consequence, lower temperatures are necessary to achieve the ground state for mechanical oscillators with lower eigenfrequencies. 

The ground state as well as coherent states are pure quantum states, but they are conventionally regarded as `classical' since they can be described by positive Glauber-Sudarshan-P-functions. To produce a \emph{nonclassical} state, the mechanical state of motion not only needs to be cooled but also coupled, for instance, to an optical field. 
Recently, strong coupling was realized in an optomechanical system with an eigenfrequency around 80\,MHz \cite{Verhagen2012}. In another work, the quantum coupling of a light field to the motion of a membrane (at about 1.5\,MHz), with an effective mass of 7\,ng in terms of radiation pressure noise (quantum back-action), was demonstrated \cite{Purdy2013}, as well as ponderomotively generated squeezed light in similar regimes \cite{Safavi-Naeini2013,Purdy2013b}. 
If the coupling is sufficiently strong, and the initial mechanical and optical states are of high enough purity, radiation pressure can lead to entanglement between field quadratures of the reflected light and the mirror's position and momentum 
\cite{Bose1997,Bose1999,Vitali2007,Hofer2011}. Very recently, entanglement between the motion of a mechanical oscillator and a propagating microwave field was generated using an electromechanical circuit \cite{Palomaki2013}. In principle, such entanglement can be used to entangle the centre of mass motions of two mechanical oscillators using a protocol that is called `entanglement swapping' \cite{Mancini2002a,Zhang2003b,Pinard2005,Pirandola2006}, thereby creating a mechanical system that is directly analogous to the Einstein-Podolsky-Rosen gedanken experiment \cite{Einstein1935}. 

Whereas the first proposals for opto-mechanical entanglement generation considered high-finesse cavities to shield the opto-mechanical system from disturbance by the environment, M\"uller-Ebhardt \emph{et al.}~\cite{Mueller-Ebhardt2008,Mueller-Ebhardt2009} found that also an open quantum system, which is realized in a Michelson interferometer, is also a suitable configuration to entangle the motion of mirrors. It was shown that steady-state motional entanglement can even be generated at room temperature if the information about the joint mirror motion that is carried away by the light is continuously acquired by photo-electric detection and used to define a conditional state of mirror motion. Additional requirements are a measurement precision close to the standard quantum limit  (SQL) of an external force measurement (as defined in \cite{Braginsky1995}), and thus a noise level from coupling to the environment that is below the SQL. As shown in \cite{Mueller-Ebhardt2008,Mueller-Ebhardt2009} it is in principle possible to conditionally prepare kg-sized mirrors that are suspended as pendula in entangled states with current state-of-the-art gravitational-wave detector technology. 

In this work, the generation of Einstein-Podolsky-Rosen entangled centre-of-mass  motion of two massive and pendulum-suspended mirrors is discussed, building on recent proposals mentioned above \cite{Mueller-Ebhardt2008,Mueller-Ebhardt2009,Miao2010,Miao2010a}, and a specific Michelson-interferometer-type setup is proposed.
The mirrors considered have masses of the order of 0.1\,kg, where the word `massive' is used as a distinction from just `macroscopic'. The latter is often defined to be `visible to the naked eye'.  
One might define `massive objects' to be those whose weight one can feel (${\gtrsim}$ 1 g), having eigenfrequencies in the audio-band or even below.

\section{Einstein-Podolsky-Rosen entanglement} 

The purpose of this section is to provide a condensed description of the phenomenon of EPR entanglement. 
In their famous 1935 publication \cite{Einstein1935}, Albert Einstein, Boris Podolsky, and Nathan Rosen (EPR) considered two particles in a specific quantum state of motion, which Erwin Schr\"odinger called an (EPR) `entangled' state in the same year \cite{Schroedinger1935}. 
EPR used this particular state to highlight why they believed that quantum theory was incomplete. 
Their belief originated from an apparent inherent contradiction in quantum theory.    

One of the essential features of quantum theory is Heisenberg's Uncertainty Principle \cite{Heisenberg1927}. It is well known that, according to quantum theory, one can not  simultaneously know the precise position and momentum of a physical system `$j\!$\;'. The product of the uncertainties, here given as variances,  has a lower bound given by 
\begin{equation}
\Delta^2(\hat x_j) \cdot \Delta^2(\hat p_j) \geq \hbar^2/4 \, . 
\label{eq:hup}
\end{equation}
Quantum theory also requires that \emph{no} such fundamental lower bound exist for the uncertainties of the \emph{relative} positions and momenta of two (entangled) physical systems $A$ and $B$:
\begin{equation}
\Delta^2(\hat x_A + \hat x_B) \cdot \Delta^2(\hat p_A - \hat p_B) \geq 0 \, . 
\label{eq:hup2}
\end{equation}
Mathematically, both inequalities are direct consequences of the commutation relations 
\begin{equation}
[\hat x_j,\hat p_j]=i\hbar \, ,
\label{eq:comm1}
\end{equation}
\begin{equation}
[\hat x_A + \hat x_B,\hat p_A - \hat p_B]=0 \, .
\label{eq:comm2}
\end{equation}
The second commutator can easily be derived from the first commutator, implying that the two commutators as well as the two inequalities above are fundamentally linked in quantum theory. 
Further, related commutators are $[\hat x_A - \hat x_B,\hat p_A + \hat p_B]=0$ and $[\hat x_A + \hat x_B,\hat p_A + \hat p_B]=[\hat x_A - \hat x_B,\hat p_A - \hat p_B]=2i\hbar$. 
These commutation relations define what position and momentum information about the two systems can simultaneously exist, in principle, and what cannot. 

The (apparent) contradiction in quantum theory arises in the following way.  
Eq.~(\ref{eq:comm2}) implies that the sum of positions and the difference of momenta of two systems $A$ and $B$ can simultaneously be defined with arbitrary precision.
This vanishing commutator thus allows for ensembles of pairs where the individual quantities $\hat x_A$ and $-\hat x_B$, as well as  $\hat p_A$ and $\hat p_B$, always have the same eigenvalues. In such ensembles the two components of a pair simultaneously carry some precise position and momentum definition \emph{with respect to each other}. 
From the environment's point of view, this allows predicting with certainty the result of a measurement on $A$ due to a measurement on $B$, without disturbing $A$. EPR argued that this feature is a sufficient condition for the `reality' of system $A$'s physical property, i.e.~for the possibility of precisely defining this observable \emph{with respect to the environment} before the measurement. This possibility would simultaneously exist for position \emph{and} momentum, and thus would contradict Eq.~(\ref{eq:comm1}), which generally implies that the position and momentum of any individual system $A$ or $B$ do not have simultaneous reality. This contradiction is also known as the `EPR paradox' \cite{Bell1964}.  

To solve this paradox, EPR conjectured that the wavefunction, as defined by quantum theory, does not provide the full information.
This led to a discussion of whether hidden variables, locally assigned to any system, existed that needed to be included in a complete theory of quantum mechanics. 
The experimentally observed violation of Bell's inequality \cite{Bell1964,Aspect1981,Giustina2013}, however, ruled out the existence of (local) hidden variables. 
Based on that, we have to conclude that the assumption made in the introduction of the paper by EPR is incorrect. Contrary to what EPR assumed, it \emph{is} in fact possible to predict the value of an arbitrary observable of a physical system $A$ with certainty via a measurement on system $B$, although this observable 
was not defined before the measurement. 
\begin{figure}[ht]
  \center
  \includegraphics[width=8.8cm]{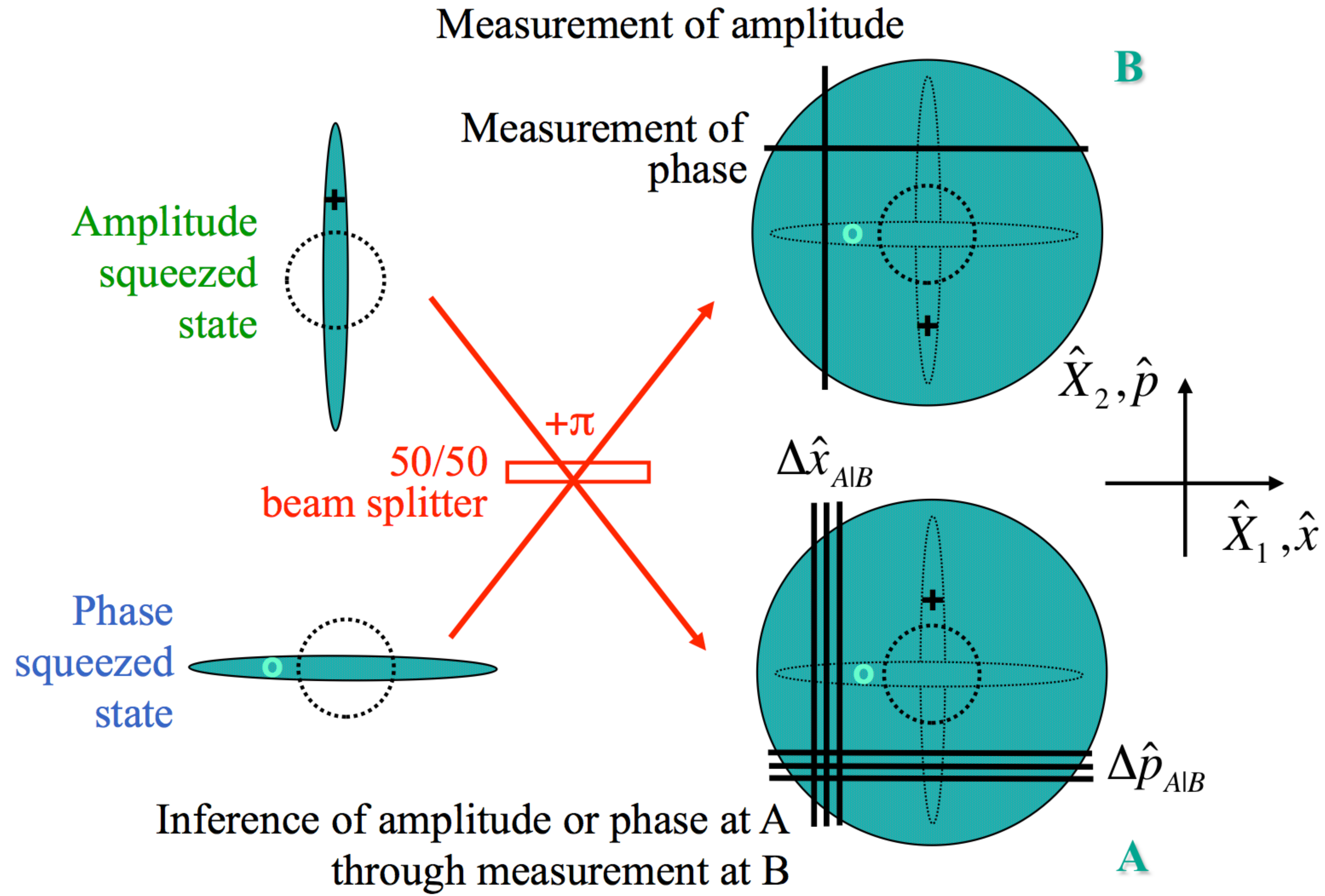}
  \caption{(color online). Illustration of the generation of (non-maximally) EPR-entangled light fields from two (non-maximally) squeezed input fields.  The ellipses and the circles display simplified sketches of quantum uncertainties in terms of cuts through the quasi-probability distributions (Wigner functions) in phase space 
{spanned by the amplitude quadrature $\hat X_1$ ($\sim \! \hat x$) and the phase quadrature $\hat X_2$ ($\sim \! \hat p$)}. 
The dashed circle represents the vacuum (ground-state) uncertainty. When two squeezed input fields (here without coherent displacement) are superposed on a (balanced) beam splitter, the two output fields individually show a rather large thermal uncertainty area, and are entangled \cite{Reid1989,Ou1992,Furusawa1998, Bowen2003a}. 
{(The `$+ \pi$' phase flip for one of the beam splitter reflections preserves energy conservation.)}
If measurements on system $B$ allow inference of the corresponding measurement result on system $A$ with an uncertainty below the vacuum state uncertainty, the state is even EPR entangled. 
The vertical line in $B$'s phase space uncertainty represents the result of an amplitude quadrature measurement. Such a measurement implies zero information about the state's phase quadrature. Since $B$'s and $A$'s amplitude uncertainties are dominated by the phase-squeezed input, their amplitude quadrature measurements are correlated. In fact, a result at $B$ allows us to infer a result at $A$ with a remaining uncertainty that corresponds to the squeezed uncertainty of the amplitude-squeezed input. This uncertainty is indicated by the three vertical lines at $A$. The situation is similar for phase quadrature measurements, only those are anti-correlated, due to energy conservation and the consequent phase flip at the beam splitter. 
Since the inference uncertainties $\Delta \hat x_{A|B}$ and $\Delta \hat p_{A|B}$ are smaller than the vacuum uncertainty, EPR entanglement is certified \cite{Reid1989}.
For infinitely strong input squeezing the uncertainties $\Delta \hat x_{A|B}$ and $\Delta \hat p_{A|B}$ approach zero and thus resemble maximal EPR entanglement and the commutation relation $[\hat x_A + \hat x_B,\hat p_A - \hat p_B]=0$. Note that in this case the local thermal uncertainties of the systems $A$ and $B$ are of infinite size.
}
  \label{fig1:EPR}
\end{figure}

The validity of quantum theory has never been put into question by any experimental result to date. All experiments that target the quantum uncertainty, be it a Bell test or a certification of EPR entanglement, fundamentally require a statistical approach, i.e.~ensemble measurements. Experiments thus give no information about the consequences of the quantum uncertainty for a single representative of the ensemble, and speculations about these consequences cannot be grounded in direct physical observations. Nevertheless, it is interesting to interpret quantum physics regarding these consequences and to check whether convincing interpretations can be developed. 
The insight above --  the fact that the possibility of predicting the value of an observable of $A$ with certainty via a measurement on $B$ 
does not require this observable `to have reality' before the measurement -- may lead to the following picture: 
first, let's take the process of spontaneous photon pair production as an example for entanglement generation. Spontaneous parametric down-conversion is frequently used to produce entangled photon pairs \cite{Burnham1970,Hong1987,Rarity1990,Kwiat1995}. 
Generally, a `spontaneous' process does not have any \emph{reason}, i.e.~it is not triggered or influenced at all by the environment.  
(It is also correct to say that the `reason' for a spontaneous process is the vacuum uncertainty that itself is fundamentally random.) 
Consequently, before the first interaction of the new photons with the environment, they are not defined from the environment's point of view (and the environment is not defined from the photon pair's point of view). From this perspective it becomes conclusive that two spontaneously generated photons can not have properties, such as polarizations, that are defined with respect to the environment, before a first interaction defines these properties. But the photons' polarizations may still be defined with respect to each other, 
{and the first interaction with the environment of one of the photons, i.e.~the first `measurement', defines the corresponding property of both.} 
Let's now consider an entangled state of motion. The entangled systems do not have individual positions or momenta that are defined with respect to the environment before the first interaction, but their positions and momenta may be defined mutually, i.e.~with respect to each other. The basis for such mutual correlations can be due to a joint origin, as it is the case for spontaneous pair production, or due to a force that acted between the systems in the past, as it is envisaged for entangling the motion of heavy objects. 
In any case, the first interaction with the environment destroys the entanglement and provides information about the position (or momentum) of one of the systems, and instantaneously also information about the second system.  From this picture it may be concluded that maximally entangled systems do not have respective physical properties such as individual positions or individual momenta at all. For systems that are less than maximally entangled, this statement needs to be weakened. In this case the uncertainties/variances in the measurement results of their individual properties describe a \emph{finite} phase space region within which a definition of the observables does not exist. 
Abandoning `reality' before the first interaction resolves the EPR paradox described above. 

EPR entanglement was observed with position and momentum-like continuous variables in a variety of experiments 
\cite{Ou1992,Hagley1997,Furusawa1998,Rowe2001,Julsgaard2001a,Bowen2003a,Bowen2003b,Blinov2004,Eberle2013}. 
Fig.~\ref{fig1:EPR} illustrates the EPR entanglement in terms of continuous (Gaussian) variables in optical experiments \cite{Ou1992,Furusawa1998,Bowen2003,Eberle2013}. As quantified by M.~Reid~\cite{Reid1989}, a necessary condition for EPR entanglement is that at least one quadrature field of one system can be inferred through a measurement of the same quadrature at the other system, with a precision that is smaller than the single-system ground state uncertainty. The EPR paradox can also be presented as Schr\"odinger's `steering effect' \cite{Schroedinger1935} as discussed in \cite{Wiseman2007} and graphically described in \cite{Haendchen2012}. 
EPR-entangled states having Gaussian quantum statistics are also called `two-mode-squeezed states'. Their Wigner function is positive, but they belong to the class of `nonclassical states' because their P-function \cite{Sudarshan1963,Glauber1963} can not be interpreted as a classical mixture of coherent states \cite{Gerry2005,Haendchen2012}.

\section{EPR-entangled motion of two massive objects}

This section discusses the realization of an EPR-entangled state of motion of two closely located pendulum-suspended mirrors of 0.1\,kg, building on earlier work in the context of gravitational-wave detectors~\cite{Mueller-Ebhardt2008,Mueller-Ebhardt2009}.
The physical concepts for the generation of the EPR entangled motion of these two mirrors are described, as well as requirements on classical noise reduction, and entanglement verification.

\subsection{EPR-entangled mirror motion in a Michelson interferometer}
\begin{figure}[ht]
  \center
  \includegraphics[width=8.6cm]{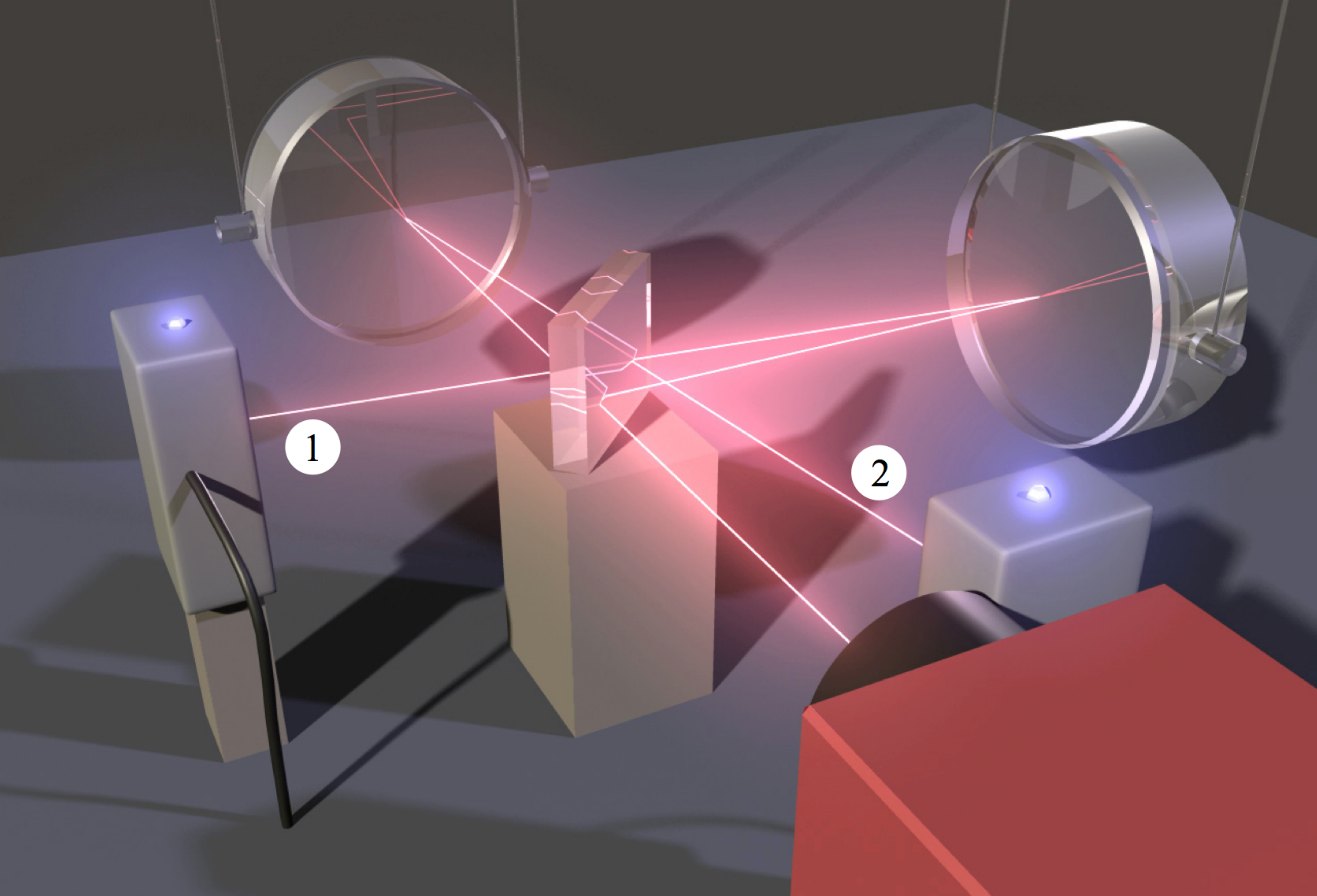}
  \caption{(color online). Illustration of the Michelson interferometer setup for the generation of an EPR entangled motion of two massive mirrors that are suspended as pendula. If the mirrors are well isolated from the environment, radiation pressure leads to entanglement between the motion of the mirror and the field quadratures of the reflected light field in each arm. Superimposing the two light fields and balanced homodyne detection at the two output ports (1, 2) enable entanglement swapping that continuously generates an EPR entangled motion of the mirrors. Residual coupling to the environment continuously destroys the entanglement, such that the entanglement will be present continuously but only over some finite short time interval. To achieve the required state purity at nonzero temperatures, conditional states of motion need to be defined (see main text) -- Courtesy of Alexander Franzen.   
\vspace{2mm}   }
  \label{fig2:MichelsonSketch}
\end{figure}

%
Let us consider a Michelson interferometer having two end mirrors that are individually suspended as pendula, as shown in Fig.~\ref{fig2:MichelsonSketch}. For the sake of simplicity, let the pendulum motions initially be in their quantum ground states. Monochromatic, continuous-wave laser light in a coherent state is injected into one input port, splits at a balanced beam splitter, travels along the arms, is retro-reflected and perfectly overlaps at the beam splitter. The differential arm length is such that all light  is retro-reflected towards the light source, while the second interferometer output port is thus at a dark fringe. The quantum states of the electro-magnetic fields leaving the two interferometer ports are sent to balanced homodyne detectors performing ideal measurements of one field quadrature at a time. 
The measurement at the dark port provides information about the differential mode of mirror motion; the one at the bright port provides information about the common mode of mirror motion. 
The homodyne angles constitute parameters to optimize the information transfer about either the joint mirror position or the joint mirror momentum, or a linear combination of the two. 
Generally, any such measurement provides information about both mechanical observables. The momentum is deduced from the difference of two subsequent position values. 
Note, that a Michelson interferometer is the natural setup for the generation of EPR entangled mirror motion, since the two output ports inherently provide information about the two joined quantities in inequality~(\ref{eq:hup2}). 

Due to reflection of the light, neither the motional quantum states of the pendula nor the light beams remain in their respective states, and instead the optical fields and the mechanical mirror motion become coupled, in the following way. 
When light in one interferometer arm is reflected from the mirror, its phase at the beam splitter is first influenced by the position of the mirror. In addition, 
the time derivative of the phase at the beam splitter is influenced by the momentum of the mirror. (Note, that a change of the light's amplitude quadrature due to mirror momentum can be neglected, since the mirror speed is always much smaller than the speed of light.)  
Next, the reflection transfers momentum from the light to the mirror, which is called radiation pressure and which leads to an excitation of the motional mirror states, affecting both its position and momentum. 

Since the light's amplitude and phase quadratures, as well as the mirror's position and momentum, have quantum uncertainties, the coupling leads to entanglement between the reflected light and the mirror motion, which can be observed if the state's purity is sufficiently high \cite{Miao2010b}. 
Finally, the entanglement of the two light/mirror-systems has to be `swapped' to entanglement of the two mirrors, by detecting the interference results of the two light fields leaving the interferometer with balanced homodyne detectors  \cite{Pirandola2006}. 
This detection simply provides information about how the initial systems are related to each other. Entanglement is not destroyed because no measurement on individual systems is performed. For instance, one detector collects information about the differential mechanical phase quadrature, and the other detector collects information about the sum of the mechanical amplitude quadratures. Both measurements are continuously performed using balanced homodyne detectors, and the recorded data streams are part of the verification process of the mirror entanglement. 

The simple Michelson topology with two balanced homodyne detectors as shown in Fig.~\ref{fig2:MichelsonSketch} is in principle sufficient for the generation of entangled mirror motion. In practice, however, using the simple Michelson interferometer is too demanding to provide a sufficiently low decoherence level. 
Two different proposals were made to reduce the information loss to the environment. The first approach is putting an optical cavity with a long lifetime around the optomechanical system \cite{Bose1997,Pirandola2003a,Vitali2007}. The second approach is making sure that the majority of the information that leaves the optomechanical system is accessible and measured with appropriate detectors \cite{Mueller-Ebhardt2008,Mueller-Ebhardt2009}. Similar to the procedure of entanglement swapping, this data stream then needs to be incorporated in the process of defining an optomechanical state, that is of high purity when `conditioned' on this data (see further down). 

This work uses the second approach. In order to increase the photon radiation pressure coupling to the heavy mirrors, optical cavities are also incorporated in the Michelson interferometer. But here the lifetimes of the cavities are rather short to avoid a reduction of the information flow from the optomechanical system required for conditioning. In the system model the cavities can thus be adiabatically eliminated \cite{Mueller-Ebhardt2008}.
As already used in the gravitational wave detector GEO\,600, retro-reflecting cavity mirrors in the two interferometer output ports are suitable to realize resonant light-power and signal enhancements, so-called `power-recycling' and `signal-recycling' \cite{Drever1983,Meers1988}.  
\begin{figure}[ht]
  \center
  \includegraphics[width=8.0cm]{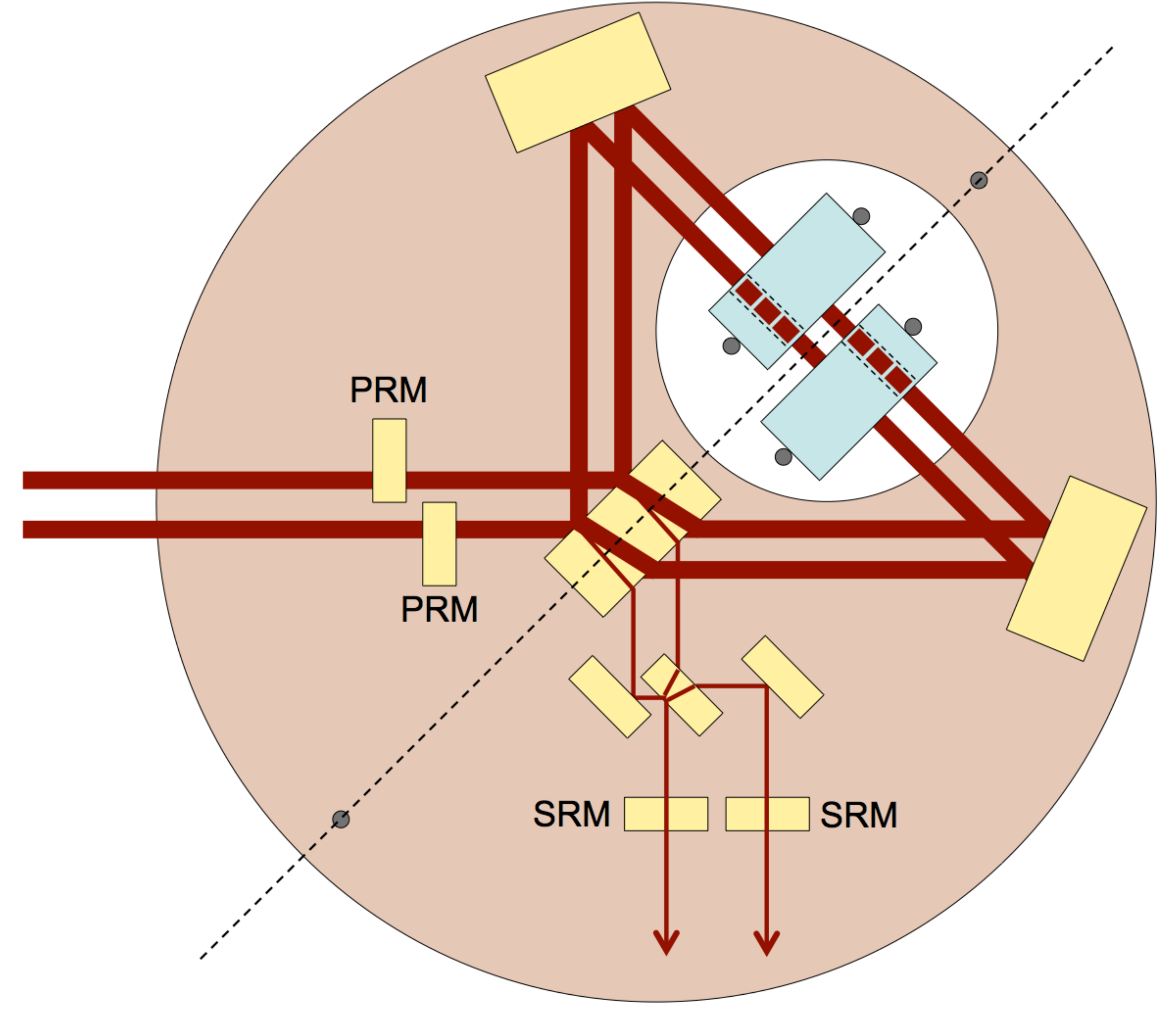}
  \caption{(color online). Proposed Michelson-type setup for EPR entanglement generation. Two test mass mirrors are suspended using fibres (small circles), as is the platform that supports all other components. Each test mass is used as the joined end mirror of a Michelson interferometer. This configuration balances the light beams' DC radiation pressure forces, which reduces the transient response when the laser is switched on.  
 The two mirrors are located close to each other, which improves the setup's sensitivity to gravitational induced state reduction \cite{Diosi1987,Diosi1989,Penrose1996, Diosi1998,Miao2010}. The dashed light beams inside the mirrors depict light paths that are guided through bore holes. 
Altogether, there are still just two optical readouts, which can both be operated close to an interferometer dark fringe, to reduce the light power on the photo diodes. The mirrors have at least one convex surface, are coated on both sides and are operated as off-resonant cavities, which reduces the number of coating layers whose thermally excited motion is directly sensed by the light beams \cite{Nakagawa2002}. Changing the laser frequency and/or the mirror temperature allows one to change the mirror transmission. Non-zero transmission is helpful to excite a Sagnac mode for aligning the setup \cite{Friedrich2011a,Westphal2012}. 
Stabilizing the mirror etalons to off-resonance might be realized by Pound-Drever-Hall locking \cite{Drever1983} and a phase modulation frequency at about half the etalons' free spectral range. In general, the mirror etalons will have different lengths, but for reasonable finesse values the off-resonance conditions should always be simultaneously achievable. 
To increase the opto-mechanical coupling, power- and signal-recycling mirrors (PRM, SRM) are used to establish cavities \cite{Drever1983,Meers1988}. The cavities have rather large bandwidths, such that information about the optomechanical state can still be efficiently detected at the balanced homodyne detectors for entanglement swapping and state conditioning.}
  \label{fig3:Michelson2}
\end{figure}

{\subsection{Proposal of an interferometer topology to increase the feasibility of mirror entanglement}} 

{This work proposes a new interferometer topology for the generation of mirror entanglement, as shown in Fig.~\ref{fig3:Michelson2}. Each mirror constitutes the common end mirror of its own Michelson interferometer. The two dark port  output fields of the interferometers are overlapped at an additional balanced beam splitter and the two resulting fields sent to two balanced homodyne detectors. 
First of all, a feature of the proposed topology is that the two mirrors are placed close to each other, face to face, which might be essential for yet-to-be-defined tests of macroscopic quantum mechanics. This way, a system is realized in which the one-dimensional motions point along a common axis. In order to achieve a stable face to face operation point, the proposed topology offers practical advantages compared to the version of the Michelson interferometer that only has folded arms. Since each mirror is the common end mirror of a Michelson interferometer, the classical radiation pressure fluctuations, which are due to a fluctuating light power and which accelerate the mirrors' centers of mass, cancel out. Similarly, the effective DC radiation pressure force on the two mirrors also cancels out, which facilitates reaching the operational point of the interferometer when switching on the laser. Perfect cancellation requires that both beams hit the mirror at precisely opposite sides. Experimentally this is achieved by using light that is transmitted through the mirror and by maximizing the interference contrast between transmitted and reflected light \cite{Friedrich2011a,Westphal2012}. More information about the proposed mirror properties is given in the caption of Fig.~\ref{fig3:Michelson2}. Another important practical advantage of the new topology is that both output fields are `dark', i.e., in case of perfect interference contrast, completely free from any carrier light and thus free from laser noise. (In reality, the contrast will not be perfect; the interferometers will also have unequal arm lengths, and it will be necessary to stabilize laser light power and frequency to some degree.) 
Furthermore, since the output fields are dim, they can be easily tomographically analysed by conventional balanced homodyne detectors. In contrast, the analysis of \emph{bright} fields with such detectors is demanding since they require a local oscillator power that exceeds the signal beam power by much more than an order of magnitude; and homodyne detectors use PIN photodiodes, which can only handle powers of up to 100 mW at best.
}
\begin{figure}[h!!!]
  \center
  \includegraphics[width=6.4cm]{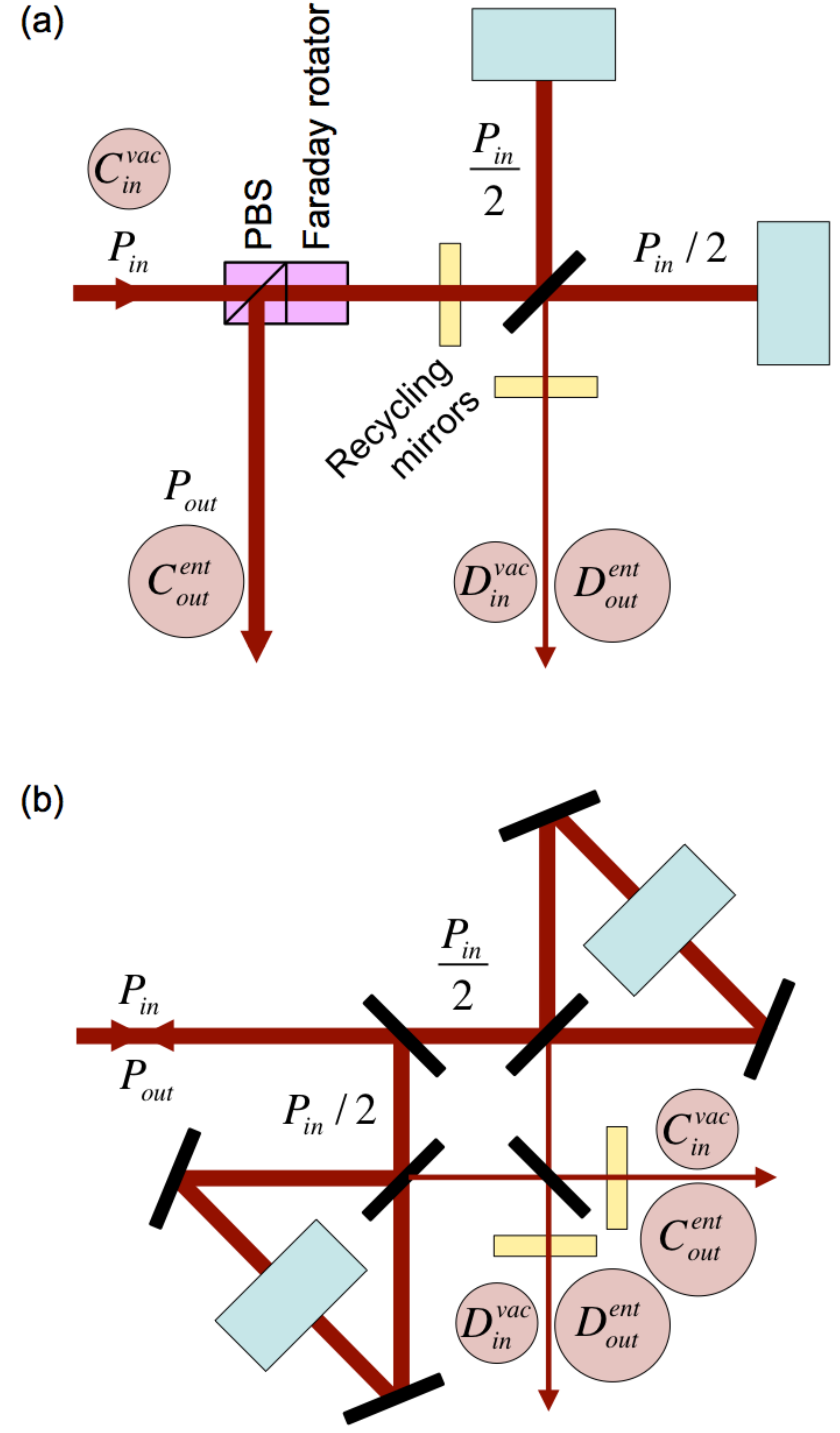}
  \caption{(color online). Equivalence of the quantum noise ports in the simple cavity-enhanced Michelson interferometer (a) and the (unfolded) topology proposed here (b). In both cases, the optical path lengths are controlled such that light beams represented by thin lines do not contain carrier light. The circles represent uncertainties of the optical states in phase space. Label \emph{C} represents the uncertainty of the mode that is relevant for the measurement of the \emph{common} mode of mirror motion. This uncertainty beats with the carrier light (of total power $P_{in}$) in such a way that the quantum noise is correlated in both interferometer arms. Label \emph{D} represents the uncertainty of the mode that is relevant for the measurement of the \emph{differential} mode of mirror motion. It beats with the carrier light in such a way that the quantum noise is anti-correlated in both interferometer arms.  
  The smaller circles represent vacuum states. The slightly larger circles indicate entanglement with the common or differential modes of mirror motion, respectively. Black bars either represent high-reflectivity mirrors or balanced beam splitters. PBS: polarising beam splitter; in combination with the Faraday rotator it removes the spatial degeneracy of input and retro-reflected light in case of the simple Michelson interferometer. Both setups have further ports, but they are not relevant in terms of the light's quantum noise. (a): Further modes that enter the PBS are either orthogonally polarised or leave the setup towards the left. (b): Quantum noise that enters the open port of the first beam splitter is eventually retro-reflected or reflected towards the left.
  }
  \label{fig4:Topology2}
\end{figure}

Another interesting feature of the proposed topology is that, in general,  it not only allows for measurements on the \emph{joint} but also on the \emph{individual} mirror motions. In this case the final beam splitter in front of the signal recycling mirrors in Fig.~\ref{fig3:Michelson2} needs to be bypassed. 
Further down an interferometric 
method to realize fast switching between the bypass and no-bypass conditions is discussed. In contrast to previously proposed entanglement verification schemes, the possibility of individual mirror quantum state tomography might open a straight-forward way of verifying motional entanglement.

Although the new interferometer topology is quite different from a single, cavity-enhanced Michelson interferometer considered in previous works, its photon radiation pressure noise is described by the same equation, and the observables measured by the two balanced homodyne detectors are not changed. In particular, the number of effective open ports through which modes in vacuum states enter the interferometer does not change. Fig.~\ref{fig4:Topology2} illustrates the equivalence of the single Michelson interferometer and the topology of the proposed topology in terms of quantum measurement noise and quantum back-action noise, which are both given by the overall light power used to sense the mirror motions, and the relevant ports of the topology. \\

\subsection{High-purity motional states conditioned on information collected from continuous measurements in the past} 
%
The observation of distinct quantum effects usually requires small couplings to the environment, i.e.~systems in rather pure quantum states. 
The usual approach to generate (unconditionally) pure mechanical and optomechanical states uses high-quality mechanical oscillators in combination with low-linewidth optical cavities. In past years, optical cooling via anti-Stokes scattering of light  
\cite{Braginsky1967,Blair1995,Gigan2006,Kleckner2006,Arcizet2006,Schliesser2006} was investigated as a useful tool for reaching the motional ground state of mechanical oscillators \cite{Teufel2011,Chan2011}. 
It requires  the sideband-resolved regime \cite{Wilson-Rae2007}, i.e.~a cavity linewidth much smaller than the mechanical frequency.  
Furthermore, the generation of unconditionally pure \emph{nonclassical} states of motion requires an optomechanical decay rate smaller than the optomechanical coupling rate, the so-called strong coupling regime \cite{Verhagen2012}. For massive mirrors, sideband-resolved optical cooling as well as reaching the strong coupling regime are unfeasible, because of the low mechanical eigenfrequencies and the low optomechanical coupling rate. The solution is the \emph{conditional} generation of high-purity optomechanical states. The mathematical description of this approach was presented in  \cite{Mueller-Ebhardt2008,Mueller-Ebhardt2009,Miao2010} and reviewed in \cite{Chen2013}.
\begin{figure}[ht]
  \center
  \vspace{2mm}
 \includegraphics[width=8.5cm]{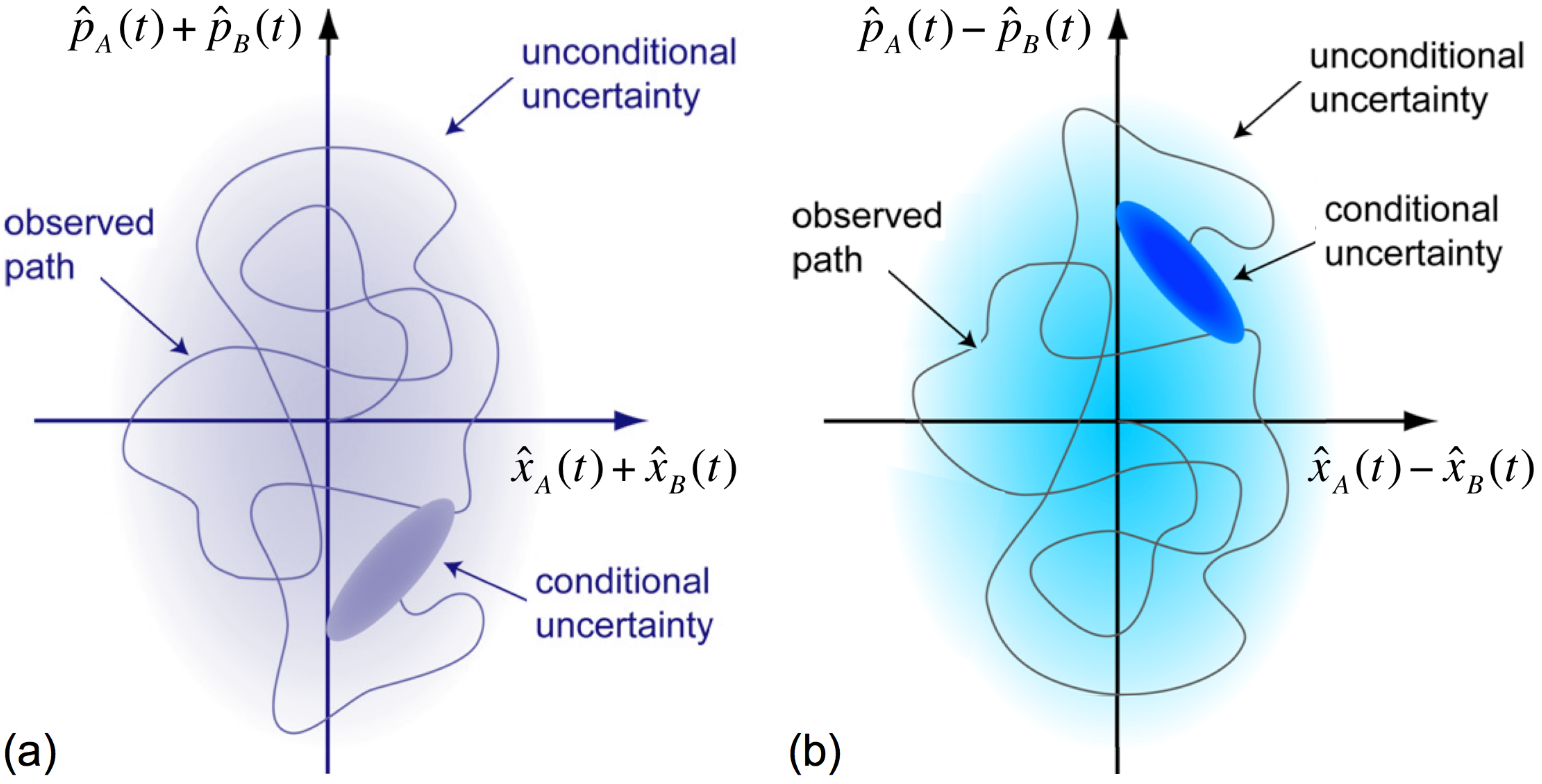}
  \caption{(color online). Phase space uncertainties of the common (a) and differential (b) mirror motion. The full shaded areas represent the highly mixed unconditional states, which are connected to the thermal bath and driven by unwanted fluctuating driving forces acting on the mirrors. The darker and smaller ellipses represent the pure conditional motional states. Conditioning on information collected from continuous measurements in the past, which provide information about the joint position and momentum, effectively `removes'  the random walk produced by stochastic excitations. The conditional states can be almost pure, depending on the precision of the information collected. The classical noise needs to be below the standard quantum limit over some band of the Fourier spectrum. The common and differential mirror motions are projected into squeezed states. If their phase-space orientations are different, the overall conditional state corresponds to the optical analogy in Fig.~\ref{fig1:EPR}, and the two mirrors are prepared in an (conditional) EPR entangled state \cite{Mueller-Ebhardt2008,Mueller-Ebhardt2009}.  
}
  \label{fig5:conditioning}
\end{figure}

Using the approach of conditional generation of pure mechanical states,  displaced minimum uncertainty states (coherent states)  as well as highly pure nonclassical mechanical states can be generated via optomechanical interaction outside the sideband-resolved and strong coupling regimes. The necessary requirement is that the information that leaves the optomechanical system is not lost to the thermal bath but collected via highly precise photo-electric detection. 
Let's consider again the two balanced homodyne detectors in the output ports of the interferometer. The detected quadrature amplitudes of the output light contain information about the mirror motions.   
The dark port detector might be optimized to collect information about the differential mirror \emph{momentum}  $(\hat p_A - \hat p_B)$, and the bright port detector to collect information about the sum of the \emph{positions} $(\hat x_A + \hat x_B)$. The distinction of whether momentum or position is efficiently collected is made by the measurement strength. The optimum measurement strategy for momentum requires less back-action, which can for example be realized by detecting an optical quadrature angle different from the usual phase quadrature.  The continuous measurements not only perform the entanglement swapping as discussed earlier, but also allow for reconstructing the classical phase-space trajectories of the joint mirror motion -- if a precise noise model for the setup allows one to optimally estimate mirror motion using Wiener filtering (in case of stationary colored noise) or Kalman filtering (in case of white noise) \cite{Mueller-Ebhardt2009}. The motional states are `pure' if conditioned on this classical phase-space information. 
Such states are also called `posteriori' states. 
The mechanical quadrature amplitudes for which the most precise information are collected become `squeezed'.  
The overall motional state shows two-mode squeezing and thus EPR entanglement.  
The measurement frequency, which is directly connected to the measurement precision  \cite{Mueller-Ebhardt2009,Chen2013}, needs to be higher than the coupling rate to the thermal bath. As shown in \cite{Mueller-Ebhardt2008}, this requirement is met if classical noise sources including thermal noise are below the spectral density of the standard quantum limit over some finite Fourier frequency band. 
Fig.~\ref{fig5:conditioning} visualizes the effect of such conditional state preparation. \\

\subsection{Noise requirements for the proposed setup and the standard quantum limit}
%
{
The main purpose of the new interferometer topology proposed here 
is to make the generation of EPR-entangled motion of two heavy mirrors feasible within a small setup. This section presents a quantitative analysis of the noise requirements. 
As it was shown in  \cite{Mueller-Ebhardt2008}, force noise acting on the masses as well as sensing noise need to be sufficiently low, whereby the standard quantum limit (SQL) sets an important reference level.
}
The SQL is a consequence of Heisenberg's uncertainty relation and refers to the sensitivity that can be achieved in the continuous measurement of an observable that does not commute with itself at different times. It constitutes an important reference for the sensitivity of a laser interferometer whose purpose is the measurement of an external force through the displacement $x$ of its mirrors. Since the susceptibility of the mirrors to the external force depends on the Fourier frequency (sideband frequency), the SQL is usually given as a (one-sided) power spectral density, normalized to $\rm{m^2}/\rm{Hz}$. 
A laser interferometer is said to be at the SQL at sideband angular frequency $\Omega = 2 \pi f >\!0$, 
if, for zero external force, the spectral density is solely due to the quantum noise of the circulating light that is in a minimum uncertainty Gaussian state, and measurement noise and back-action noise are uncorrelated as well as balanced \cite{Braginsky1968,Caves1980,Caves1981} (see also the short discussion in the introduction of Ref.~\cite{Westphal2012}). 
If the susceptibility of the mirrors correspond to that of a free mass, the (`free-mass') SQL reads \cite{Braginsky1968,Braginsky1995}
\begin{equation}
S_x^{\rm{fmSQL}}(\Omega)= \frac{2 \hbar}{m \Omega^2}  \, .
\label{eq:SQL}
\end{equation}
where  $m$ is the mass, whose displacement $x$ probes external forces. 
In a simple Michelson interferometer (Fig.~\ref{fig2:MichelsonSketch}) as well as in the setup shown in Fig.~\ref{fig3:Michelson2}, $m$
{corresponds to the reduced mass $m = m_0/2$, where $m_0$ is the mass of one mirror, and $x$ corresponds to the differential displacement $x_A - x_B$.} 
If the susceptibility of the mirrors corresponds to that of a harmonic oscillator, the (`harmonic oscillator') SQL reads
\begin{equation}
S_x^{\rm{hoSQL}}(\Omega)= \frac{2 \hbar}{m | - \Omega^2 +  \Omega_0^2 + i  \frac{\Omega \Omega_0}{Q}|}  \; ,
\label{eq:hoSQL}
\end{equation}
where  $\Omega_0$ is the oscillator's resonance frequency and $Q$ its quality factor.

As shown in Refs.~\cite{Mueller-Ebhardt2008,Mueller-Ebhardt2009}, the SQL represents an important reference for the possibility of generating conditional EPR-entanglement of the motion of two mirrors. It can be generated in the presence of excess noise (classical noise) -- but only if the sum of all classical noise spectral densities is below the SQL over some finite frequency band. 
\begin{figure}[ht]
  \center
 \includegraphics[width=8.6cm]{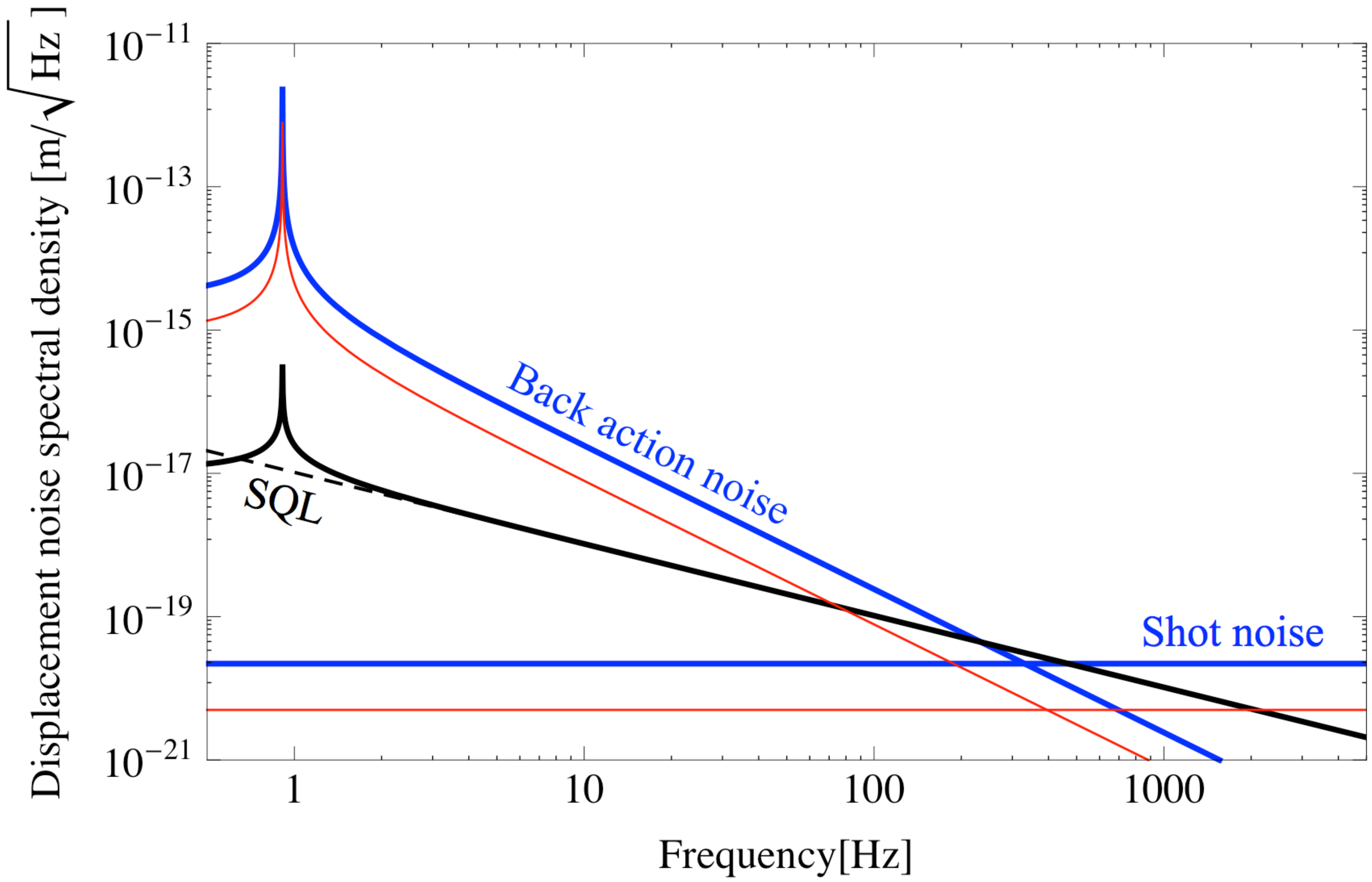}
  \caption{(color online). Example noise spectral densities of the interferometer shown in Fig.~\ref{fig3:Michelson2}, normalized to differential mirror displacement \cite{Buonanno2003b,Yamamoto2010} for mirror masses of 0.1\,kg being suspended from 30\,cm long fibers.
Back-action noise and shot noise are calculated for a light power of 1\,kW in each interferometer arm, and for a signal-recycling gain of 2000 in power. The standard quantum limit (SQL) is given for the actual oscillator (solid line) as well as for the corresponding free-mass system (dashed).
The peaked curve directly below the curve `back action noise' shows the pendulum thermal noise spectrum at 4\,K. 
The horizontal line below the `shot noise' shows the targeted level of the total `classical sensing noise'  \cite{Mueller-Ebhardt2009} being dominated by coating thermal noise. This level might be achievable at 4\,K (see main text). 
In the example shown, the SQL is achieved at about 300\,Hz. Pendulum thermal noise as well as classical sensing noise are below the SQL from about 80 to 2000\,Hz, providing a large margin for the generation of conditional motional EPR entanglement \cite{Mueller-Ebhardt2008,Mueller-Ebhardt2009}. Note that the \emph{momentum} readout requires less back-action and the SQL needs to be achieved at a slightly lower frequency \cite{Mueller-Ebhardt2008}, which can be achieved by a lower signal-recycling gain for the momentum read out in Fig.~\ref{fig3:Michelson2}  or by detecting the output field using a quadrature angle that does not correspond to the intra-interferometer phase quadrature.} 
  \label{fig6:spectrum}
\end{figure}

Fig.~\ref{fig6:spectrum} shows the square root of the free-mass SQL (dashed) for the setup in Fig.~\ref{fig3:Michelson2} with mirrors of mass $m_0 = 0.1$\,kg, as well as the oscillator SQL when the mirrors are suspended as pendulums of length $L = 30$\,cm with a mechanical quality 
{of $Q  = 2\!\cdot\!10^7$,} as for instance realized in \cite{Rowan1997}.

Fig.~\ref{fig6:spectrum} also shows the square roots of the setup's quantum measurement (shot noise) and back-action spectral densities for a total light power of $P\!=\!4$\,kW (1\,kW on each of the four mirror surfaces), and a signal recycling gain of $G_{\rm{SR}} \approx 4 / t_{\rm{SR}}^2 = 2000$, where $1 - t_{\rm{SR}}^2 = 99.8\%$ is the signal recycling mirror's reflectivity. The total circulating light power might be generated using an input power of 2\,W and a power-recycling mirror reflectivity of $99.8\%$ as well. 
The spectral density is calculated following the work in Refs.~\cite{Buonanno2003b,Yamamoto2010}. Approximated for sideband frequencies much smaller than the signal-recycling cavity linewidth, it reads
\begin{equation}
S_x^{\rm{SN}} = e^{-2r} \frac{\hbar c^2}{2 \omega P G_{\rm{SR}}}  \, .
\label{eq:SN}
\end{equation}
Here, $r$ is the squeezing parameter \cite{Gerry2005} of pure squeezed vacuum states that are injected into the signal output ports, mode-matched to the output fields, and phase-controlled such that the shot-noise is squeezed, i.e.~reduced \cite{Schnabel2010}. Since 2010 squeezed light has been regularly used for the same purpose in the gravitational wave detector GEO\,600 \cite{LSC2011,Grote2013}, and was also successfully tested in one of the LIGO detectors \cite{Aasi2013}. For a squeezing strength of 10\,dB ($e^{-2r}\!=\!0.1$) \cite{Vahlbruch2008,Eberle2010,Mehmet2011}, the light power requirement reduces by a factor of ten in order to achieve the SQL (at a given sideband frequency).

The quantum back-action noise, again approximated for small sideband frequencies, then reads \cite{Buonanno2003b,Yamamoto2010} 
\begin{equation}
S_x^{\rm{RPN}}(\Omega)= \frac{e^{2r}}{m^2 | - \Omega^2 +  \Omega_0^2 + i  \frac{\Omega \Omega_0}{Q} |^2}   \frac{2 \hbar \omega P G_{\rm{SR}}}{c^2}  \, .
\label{eq:RPN}
\end{equation}

The peaked curve directly below the `back action noise' in Fig.~\ref{fig6:spectrum} represents the thermal noise spectral density of the pendulum for viscous damping \cite{Gardiner2004}, 
and for a temperature of $T = 4$\,K and mechanical quality of $Q  = 2\!\cdot\!10^7$, according to
\begin{equation}
S_x^{\rm{Pen}}(\Omega)= \frac{1}{m^2 | - \Omega^2 +  \Omega_0^2 + i  \frac{\Omega \Omega_0}{Q} |^2}  \frac{4\, m  \Omega_0 k_B T}{Q}  \, .
\label{eq:RPN}
\end{equation}

It represents `classical force noise' \cite{Mueller-Ebhardt2009} that results in decoherence of the mirror's centre of mass motion, and that needs to be as low as possible. Again, the equation provides the one-sided spectrum, $m$ is the reduced mass, and $x$ corresponds to the differential displacement. Classical power fluctuations of the laser light as well as seismic noise further contribute to force noise. Here, it is assumed the latter can be made insignificant by modern laser stabilization techniques \cite{Kwee2012} and isolation systems \cite{Harry2010d}, respectively. 

The lowest horizontal line in Fig.~\ref{fig6:spectrum} represents the desired goal for the total `classical sensing noise' \cite{Mueller-Ebhardt2009} of about $5 \cdot 10^{-21} \rm{m}/ \sqrt{Hz}$. For simplicity a white spectrum is plotted, although its major constituents have a colored spectrum. Sensing noise is a consequence of optical loss, mirror internal thermal noise, and mirror coating thermal noise, which is expected to be the dominant contribution in this setup \cite{Levin1998}. The sensing noise level assumed here requires coating thermal noise that is about two orders of magnitude lower than the one calculated in Ref.~\cite{Corbitt2006} for a 1\,g mirror at room temperature and for a laser beam diameter of 1\,mm. The aimed reduction seems realistic. The lower temperature provides one order of magnitude reduction. The larger dimensions of the mirrors should allow for an increased laser beam diameter to about 3\,mm, providing another order of magnitude in the reduction of coating thermal noise \cite{Corbitt2006}. Recent progress in the development of crystalline mirror coatings \cite{Cole2008} and crystalline monolithic mirrors \cite{Brueckner2010} will additionally help reduce coating thermal noise. Apart from coating thermal noise, optical absorption of laser light is a crucial issue. The absorption needs to be low enough to allow for an operation around 4\,K to keep the assumptions in Fig.~\ref{fig6:spectrum} valid. Low-absorption coatings and new ways of removing the heat from the suspended mirrors need to be developed, which is also a timely challenge for the development of cryogenic gravitational wave detectors \cite{Somiya2012}. The application of squeezed vacuum states of light will help to reduce the power requirements and amount of heat deposited in the mirrors. 
If the spectral densities shown in Fig.~\ref{fig6:spectrum} can be realized, the ratio of the frequencies at which the sensing noise and the force noise intersects the SQL will be as high as $25 > 2$, where 2 is the critical value above which the generation of mirror entanglement is possible \cite{Mueller-Ebhardt2008,Mueller-Ebhardt2009}.

\subsection{Verification of the EPR entangled motion}

In the experiment proposed here, the entangled motion is continuously generated over a finite time-interval. 
After the entanglement generation stage has been completed, a period of free evolution might follow, terminated by a measurement that verifies the successful entanglement generation. Miao \emph{et al.}~proposed and analysed in detail a feasible verification stage \cite{Miao2010}. 
To realize a phase of free evolution they suggested either turning off the light, or alternatively evading its back action.  Back action is effectively evaded by (continuously) reading out the quadrature that is responsible for the radiation pressure noise, which does not contain information about the oscillators' position, and by taking this data into account during the verification stage. The verification stage, in fact, uses data from the same balanced homodyne detectors that are also used for evading back-action during free evolution and for generating the entanglement. 
Miao \emph{et al.} showed that the mechanical oscillator's position and momentum at time $t$ can be inferred from data after $t$. In particular, they showed that this inference can reach the required sub-Heisenberg accuracy if the time-dependence of the local-oscillator phase as well as the weighting of the data collected at different times are optimized. 
The result is a reconstructed quantum state of the (entangled) mechanical oscillators, which allows entanglement verification if the expectation values of the conditionally prepared states are known and if all noise sources are Markovian \cite{Miao2010}. (In an actual experiment not all noise sources will be Markovian, however, according to \cite{Miao2010}, their main findings will not change for non-Markovian noise.)
The proposed sequence of conditional entanglement generation, its free evolution, and a final measurement for verification is thus initiated and executed by changing the local oscillator phases in the balanced homodyne detectors.  
To verify successful generation of motional EPR entanglement, the full process of entanglement generation, free evolution and measurement needs to be repeated many times in order to collect statistical data. Note, that all information collected is information about the \emph{joint} mirror motion. 

The duration of the entanglement generation as well as the (conventional) decoherence time are connected to characteristic frequencies \cite{Miao2010}, that can be deduced from the quantum and thermal noise spectral densities. In particular, the initially generated entanglement will be decohered if the period of free evolution is greater than $\tau_F$, which is given by the inverse of the angular frequency at which the force noise intersects the SQL. If the requirements in Fig.~\ref{fig6:spectrum} are met, entanglement can be observed if the period of free evolution is shorter than $\tau_F \approx 1/(2 \pi \,80\, \rm{Hz}) \approx 2$\,ms. The observation of a shorter decoherence time might point towards unconventional decoherence processes. The minimal period required for entanglement generation $\tau_q$ is related to the inverse of angular frequency at which the back-action noise intersects the SQL. The data in Fig.~\ref{fig6:spectrum} thus requires $\tau_q > 1/(2 \pi \,230\, \rm{Hz}) \approx 0.7$\,ms. The optimum duration of the verification stage $\tau_V$ is  given in \cite{Miao2010}, and always fulfills $\tau_q < \tau_V < \tau_F$.

The measurements for the entanglement verification procedure that are outlined above and that were proposed and analysed in detail in \cite{Miao2010} are suitable for the interferometer topology proposed here. 
The setup in Fig.~\ref{fig3:Michelson2}, however, might also allow for the measurement of the positions and momenta of the \emph{individual} mirrors  -- if the final beam splitter were removed. Such a measurement would be in direct analogy to the EPR gedanken experiment.  
After entanglement has been generated the verification stage then requires bypassing the final beam splitter in Fig.~\ref{fig3:Michelson2} and sending the light fields directly to balanced homodyne detectors without overlapping them. 
Fast switching could, for example, be realized by overlapping the two output fields in Fig.~\ref{fig3:Michelson2} on another balanced beam splitter. If the relative phase between the two fields is switched from zero to $\pi/2$, the effect of the preceding beam splitter can be canceled out. The setup corresponds to a Mach-Zehnder interferometer that is switched between bright fringe and half fringe. Fast switching on 0.1\,ms timescales would be short enough to initiate the verification stage within the expected decoherence time. A feasible approach might be based on a small piezo-actuated mirror, which is also the approach that introduces minimal optical loss. Even though this experiment opens the possibility to measure individual mirror motions, in full analogy to the EPR gedanken experiment, the processing of the full data set in the verification stage will still follow the proposal in \cite{Miao2010}.

\section{Conclusion}

This work considers the realization of  Einstein-Podolsky-Rosen entangled centre-of-mass motion of two 0.1\,kg mirrors that are located close to each other and that are individually suspended as pendula. The observation of EPR entangled mirror motion is, first of all, confirmation of quantum mechanics in the macroscopic world. 
But furthermore, the EPR entangled system discussed in this work might also serve as an interesting probe system to test predictions of modified quantum theories that include (nonrelativistic) gravitational effects, i.e.~approaches towards a theory of `Newtonian quantum gravity'. Di{\'o}si and Penrose proposed the existence of so-called gravity decoherence  \cite{Diosi1987,Diosi1989,Penrose1996,Diosi1998}. In this idea, the proper gravitational field of the system is conjectured to introduce decoherence to the system's motional state. In particular, a single massive particle in a motional Schr\"odinger cat state is a sensitive probe to test gravity decoherence, and experiments have been proposed to realize such a cat state \cite{Marshall2003}. Although the EPR entangled state considered in this work reflects a state with Gaussian statistics and an entirely positive quasi-probability function (Wigner function), it can in principle also be used to test  the effect of gravity decoherence. Following Miao \emph{et al.}~\cite{Miao2010} the quantum entanglement vanishes within a time scale of   $\hbar$  divided by self-energy of the mass-distribution difference and would limit the maximal free evolution in the experiment discussed here to $10^{10}$\,s. 
Such long times are experimentally not feasible but Miao \emph{et al.}~found that another potential decoherence mechanism leads to much shorter decoherence times. 
They considered decoherence caused by the mutual gravitational energy among components, which would appear on a microsecond timescale in the setup discussed here. A test of such a short decoherence mechanism seems feasible if the low noise floor of Fig.~\ref{fig6:spectrum} will be achieved, however, the rationale of such a decoherence effect seems to be less intuitive than the original Di{\'o}si-Penrose gravity decoherence.

In any case, massive objects that are suspended as pendula and whose motion is quantum mechanically entangled are interesting probes of modified quantum theories. At Fourier frequencies above the pendulum eigenfrequency, they act as quasi-free falling test masses of space-time.  Most likely, further predictions of modified quantum theories will be made in future. For some of them, the system of massive EPR entangled mirrors discussed here might find application in researching routes towards a unified theory of quantum gravity. \\

\begin{acknowledgments}
This research has been supported by the ERC project `MassQ'  (grant agreement number 339897) and EU ITN `cQOM' (grant agreement number 290161). RS thanks Haixing Miao, Yanbei Chen, Kentaro Somiya, Mikhail Korobko, and Aleksandr Khalaidovski for fruitful discussions within the quantum noise working group of the LIGO Scientific Collaboration (LSC), as well as Sacha Kocsis and Torsten Franz for their invaluable comments on the manuscript. Thanks are also due to Alexander Franzen for his help with Fig. 2.
\end{acknowledgments}


\end{document}